\documentstyle[11pt,newpasp,twoside,epsf]{article}
\def\edcomment#1{\iffalse\marginpar{\raggedright\sl#1\/}\else\relax\fi}
\reversemarginpar

\begin{document}
\title{A Semi-analytic Model of Halo Dynamics}
 \author{James E.\ Taylor}
\affil{Astrophysics, Denys Wilkinson Building, Keble Road, Oxford OX1 3RH, UK}
\author{Arif Babul}
\affil{Department of Physics and Astronomy, University of Victoria, Victoria, BC V8P 1A1, Canada}

\begin{abstract}
We have developed a new semi-analytic model for the formation 
and evolution of structure on galaxy, group and cluster scales. The model 
combines merger trees with a detailed, spatially 
resolved description of the dynamical evolution of halo substructure. It 
reproduces the results of numerical simulations of mergers and halo formation
at a fraction of the computational expense. We use this model to
study the formation of the dwarf galaxies in the Local Group and 
the structural components of the Milky Way. 
We find that reionization can explain the scarcity of Local 
Group dwarfs, although the reionization epoch is constrained to be 
high -- $z_{\rm ri} \ge 12$.
We also find that disk disruption at recent times is
rare, such that many galactic disks have an old, thin component.
\end{abstract}

\section{Introduction}

Substructure in CDM halos consists of hundreds of small, dense `subhalos'. 
The dynamical evolution of these satellites can be described using the 
analytic approach of Taylor \& Babul (2001), which remains accurate to 
about 20\% over many orbits. Combining this approach with
extended Press-Schechter (EPS) merger trees, we have developed a new, 
semi-analytic (SA) model for the dynamical evolution of cluster or galaxy 
halos. This model predicts halo substructure identical to that seen in
high-resolution numerical simulations. We have used the SA model to study 
the origin of the Local Group satellites, the stellar halo, and the thick and 
thin disk components of the Milky Way.

\section{Results}

\vspace{0.1in}
{\bf {\noindent The Origin of Local Group Dwarfs:}} By accounting for the effects of 
reionization and evolving 
mass-to-light ratios, it is possible to produce luminosity functions
similar to that of the Local Group (Figure 1, left panel), provided
reionization occurs at early epochs ($z_{\rm ri} \ge 12$). 

\vspace{0.1in}
{\bf {\noindent The Origin of the Stellar Halo:}} Debris from disrupted satellites 
also forms a plausible stellar halo, with an $r^{-3.5}$ density profile.

\vspace{0.1in}
{\bf {\noindent The Age of the Thin Disk:}}
The disk of the central galaxy can be disrupted by direct collisions with 
large satellites. The model thus provides an upper limit to the age of 
stars still in disk orbits for each system. The mean ages of the model disks 
are 8--10 Gyr, similar to the age of the Milky Way's disk. 

\vspace{0.1in}
{\bf {\noindent The Origin of the Thick Disk:}}
The disk can also be heated by small collisions or indirect encounters,
without being completely disrupted. In many model systems,
minor encounters at early times produce a heated component 
similar to the thick disk of the Milky Way (Figure 1, right panel).

\begin{figure}
\plottwo{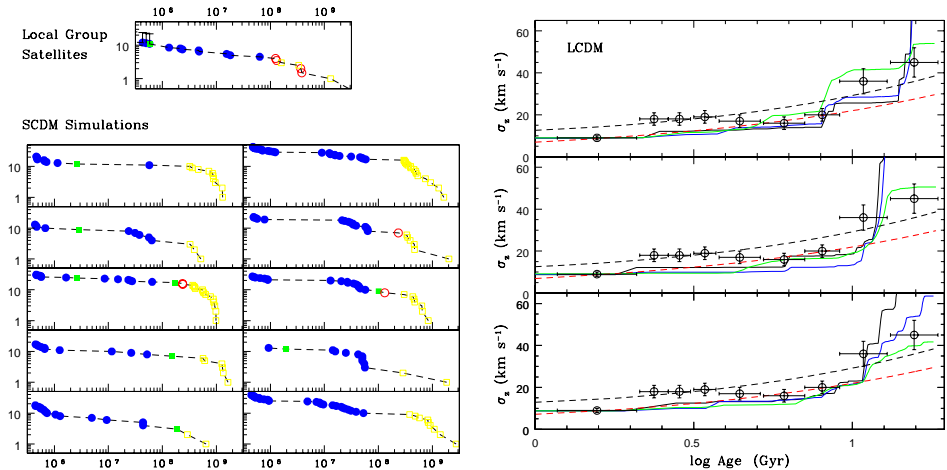}{f8.18.epsi}
\caption{Left: Typical SA luminosity functions, 
compared with that of the Local Group (top). Point types indicate
morphology.
Right: The predicted increase in velocity dispersion from 
indirect heating (solid lines), compared with observations (points, 
from Quillen \& Garnet (2001)), and smooth $t^{1/2}$ relations (dashed lines).}
\end{figure}

\end{document}